\documentclass[preprint,floatfix,preprint,showpacs,nofootinbib]{revtex4}
\usepackage{natbib}
\usepackage{times}
\usepackage{amssymb,amsbsy,amsmath,amsfonts}
\usepackage{graphicx}
\usepackage{float}
\usepackage{rotating}

\begin{document}

\preprint{\small FZJ-IKP-TH-2009-26}
\title{Study of the $f_2(1270)$, $f_2'(1525)$, $f_0(1370)$ and $f_0(1710)$ in the $J/\psi$ radiative decays}

\author{L.S. Geng,$^1$ F.K. Guo,$^2$ C. Hanhart,$^{2,3}$ R. Molina,$^1$ E.~Oset$^1$ and B.S. Zou$^{4,5}$}
\affiliation{$^1$Departamento de F\'{\i}sica Te\'orica and IFIC,
Centro Mixto Universidad de Valencia-CSIC,
Institutos de Investigaci\'on de Paterna, Aptdo. 22085, 46071 Valencia, Spain}
\affiliation{$\rm ^2$Institut f\"{u}r Kernphysik and J\"ulich Center
             for Hadron Physics,\\ Forschungszentrum J\"{u}lich,
             D--52425 J\"{u}lich, Germany}
\affiliation{$\rm ^3$Institute for Advanced Simulation,\\ Forschungszentrum J\"{u}lich,
             D--52425 J\"{u}lich, Germany}
\affiliation{$^4$Institute of High Energy Physics, CAS, P.O.Box 918(4), Beijing 100049, China}
\affiliation{$^5$Theoretical Physics Center for Science Facilities, CAS, Beijing 100049, China}

 \begin{abstract}
In this paper we present an approach to study the radiative decay modes of the
$J/\psi$ into a photon
 and one of the tensor mesons $f_2(1270)$,
 $f'_2(1525)$, as well as the scalar ones $f_0(1370)$ and $f_0(1710)$.
Especially we compare predictions that emerge from a scheme
where the states appear dynamically in the solution of  vector meson--vector meson scattering
 amplitudes to those from a (admittedly naive) quark model.
We provide evidence that it might be possible to distinguish amongst
the two scenarios, once improved data are available.
\end{abstract}


\date{\today}
\maketitle

\section{Introduction}

Interactions amongst hadrons may be sufficiently strong to produce
bound states --- this picture even emerges naturally when starting
from a quark model~\cite{extraordinary}. Famous examples of those
are nuclei, but also amongst mesons a large number of
states were recently identified as candidates for hadronic
molecules. However, only if  one channel is largely dominant and the
(quasi)--bound--state poles are located close to the corresponding
continuum threshold for the constituents that form the molecule
through an $s$--wave interaction~\cite{evidence,Weinberg:1962hj} or
if the $N_c$ behavior of the system can be controlled~\cite{rios}, a
model independent access to the nature of the state appears to be
possible. On the other hand there is a large number of proposed
molecular states, where neither of the mentioned criteria applies.
In this case it needs to be demonstrated that the molecular picture
describes the properties of the states better than, say, a
conventional quark--model description. This might emerge, e.g.,
since the pattern of SU(3)--flavor breaking turns out to be
different for hadron--hadron states, due to the different analytic
structure of their scattering amplitudes. In this paper we discuss
observables that might qualify for such a test for the possible
molecular nature of the $f_2(1270)$, $f'_2(1525)$, $f_0(1710)$, and
$f_0(1370)$.

The reactions we will focus on are decays of the $J/\psi$. The
decays of $J/\psi$ into a vector meson $\phi$ or $\omega$ and two
pseudoscalars, including their non-perturbative final state
interactions driven by the presence of the scalar mesons $f_0(600)$
and $f_0(980)$, were studied recently in
Refs.~\cite{ulfjose,palochiang,lahde}.  These processes proceed
through an OZI violating strong interaction and it was assumed that
the transition $J/\psi\to V$ provides a scalar source term allowing
for an investigation of scalar form factors (for corrections induced
by the interaction of the pseudoscalars with the vector meson see
Refs.~\cite{palochiang,liu}).
Analogous to the decay modes mentioned above are the modes $J/\psi
\to \phi (\omega) f_2(1270)$, $J/\psi \to \phi (\omega) f'_2(1525)$
and $J/\psi \to K^{*0}(890) \bar{K}^{*0}_2(1430)$.  These decays
were recently studied in Ref.~\cite{albertozou} following similar
steps as done in Refs.~\cite{ulfjose,palochiang,lahde},  within the
scheme where these tensor states are dynamically generated from the
interaction of pairs of vector mesons.  Indeed, in
Ref.~\cite{raquel} it was shown that the $f_2(1270)$ and $f_0(1370)$
states appear naturally as bound states of $\rho \rho$ using the
interaction kernel provided by the hidden gauge Lagrangians
\cite{hidden1,hidden2,hidden3,ulfvec}. An extension to SU(3) of the
former work of Ref.~\cite{raquel} done in
Ref.~\cite{gengvec,Geng:2009gb}, studying the interaction of pairs
of vectors, shows that there are as many as 11 states dynamically
generated, some of which can be associated to known resonances,
namely the $ f_2(1270),f'_2(1525)$ and $K^*_2(1430)$ resonances, as
well as the $f_0(1370)$ and $f_0(1710)$. In this paper we
investigate the same $J/\psi$ decays together with their radiative
counterparts and make predictions for ratios of decay rates for the
mentioned molecular picture and a naive quark model assignment.

\begin{figure}[b]
\begin{center}
\includegraphics[scale=0.6]{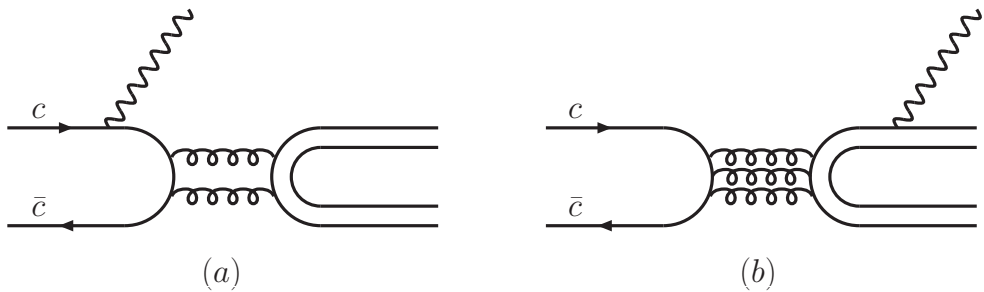}
\caption{Two mechanisms of the $J/\psi$ radiative decays.}
\label{f1}
\end{center}
\end{figure}

\section{Decay rates in the molecular picture}
Two topologies are possible for the radiative decays of the $J/\psi$
into light hadrons, as shown in Fig.~\ref{f1}. Since the photon
carries both an isospin 1 and an isospin 0 component, the hadronic
final state for the mechanism of diagram $(b)$ can have an admixture
of both isospins. In diagram $(a)$, on the other hand, also after
the photon emission, the isospin of the $\bar cc$ pair is still zero
and correspondingly the isospin of the hadronic final state is zero.
Since at the charm quark mass $\alpha_s$ is about $1/3$, relatively
small, and diagram $(b)$ involves an additional loop, diagram $(a)$
is expected to be the dominant process~\cite{physrep} and we
therefore may assume the radiative $J/\psi$ decays as a source for
isoscalar light hadrons. This is also confirmed by
data~\cite{Amsler:2008zzb}. More than this, under the dominance of
diagram (a), the state after the radiation is still a $c\bar{c}$ and
hence an SU(3) singlet, which is relevant for the present study.

\begin{figure}[htpb]
\begin{center}
\includegraphics[scale=0.6]{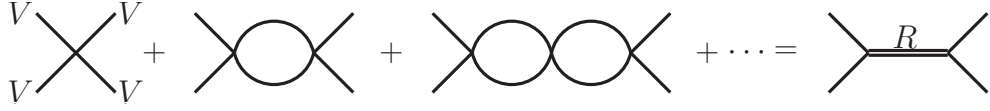}
\caption{Schematic representation of dynamically generated resonances from vector meson-vector meson interaction.} \label{f2}
\end{center}
\end{figure}
\begin{figure}[htpb]
\begin{center}
\includegraphics[scale=0.6]{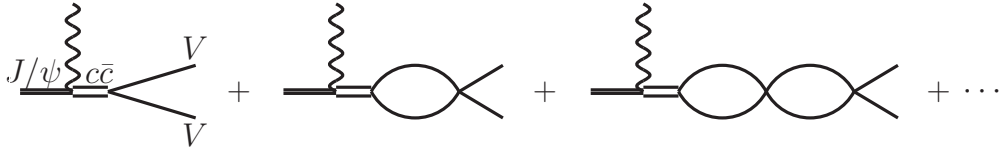}
\caption{Schematic representation of $J/\psi$ decay into a photon and one dynamically generated resonance.} \label{f3}
\end{center}
\end{figure}

We now need to calculate the formation of the
resonances. This calculation depends on the assumed
nature of the states. In this section we will discuss
the rates that emerge in a molecular picture for
the mentioned resonances --- namely they are assumed
to emerge from the non-perturbative interactions of
vector mesons amongst themselves.
 We readily
get the SU(3) singlet combination of two vectors from
\begin{equation}
 \mathrm{VV}_{\mbox{SU(3) singlet}}=\mathrm{Tr}[V.V],
\end{equation}
where $V$ is the SU(3) matrix of the vector mesons
\begin{equation}
 V=\left(\begin{array}{ccc}
          \frac{1}{\sqrt{2}}\rho^0+\frac{1}{\sqrt{2}}\omega & \rho^+ &K^{*+}\\
           \rho^- &-\frac{1}{\sqrt{2}}\rho^0+\frac{1}{\sqrt{2}}\omega & K^{*0}\\
           K^{*-} &\bar{K}^{*0}&\phi
         \end{array}
\right).
\end{equation}
We, thus, find the vertex
\begin{equation}
 \mathrm{VV}_{\mbox{SU(3) singlet}}=
\rho^0\rho^0+\rho^+\rho^-+\rho^-\rho^++\omega\omega+K^{*+}K^{*-}+K^{*0}\bar{K}^{*0}
+K^{*-}K^{*+}+\bar{K}^{*0}K^{*0}+\phi\phi.
\end{equation}
One then projects this combination over the VV  states which are the
building blocks of the resonance produced, with unitary normalization (an extra
factor $1/\sqrt{2}$ for identical particles or symmetrized ones) and
phase convention $|\rho^+\rangle=-|1,+1\rangle$, $|K^{*-}\rangle=-|1/2,-1/2\rangle$ of isospin,
\begin{eqnarray}
 |\rho\rho\rangle_\mathrm{I=0}&=&-\frac{1}{\sqrt{6}}|\rho^0\rho^0+\rho^+\rho^-+\rho^-\rho^+\rangle,\\
|K^*\bar{K}^*\rangle_\mathrm{I=0}&=&-\frac{1}{2\sqrt{2}}|K^{*+}K^{*-}+K^{*0}\bar{K}^{*0}+K^{*-}K^{*+}+\bar{K}^{*0}K^{*0}\rangle,\\
|\omega\omega\rangle_\mathrm{I=0}&=&\frac{1}{\sqrt{2}}|\omega\omega\rangle,\\
|\phi\phi\rangle_\mathrm{I=0}&=&\frac{1}{\sqrt{2}}|\phi\phi\rangle,
\end{eqnarray}
and one gets the weights for primary VV production of the process
$J/\psi\rightarrow\gamma\mathrm{VV}$:
\begin{equation}\label{eq:cg}
 w_i=\left\{\begin{array}{ll}
             -\sqrt{\frac{3}{2}}&\quad\mbox{for $\rho\rho$}\\
             -\sqrt{2}&\quad\mbox{for $K^*\bar{K}^*$}\\
             \frac{1}{\sqrt{2}} &\quad \mbox{for $\omega\omega$}\\
             \frac{1}{\sqrt{2}} &\quad\mbox{for $\phi\phi$}
            \end{array}
\right. .
\end{equation}
Note that these weights are just SU(3) symmetry
coefficients, and they are obtained with the momentum-independent
approximation of the production vertices, which is valid since the
mass differences among the vector mesons are small.

\begin{figure}[htbp]
\begin{center}
\includegraphics[scale=0.6]{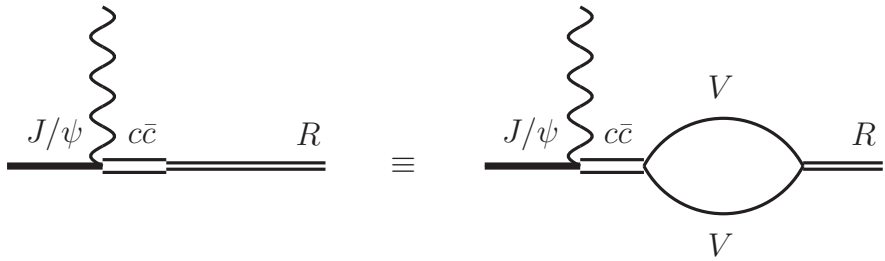}
\caption{Schematic representation of $J/\psi$ decay into a photon and one dynamically generated resonance.} \label{f4}
\end{center}
\end{figure}
The next step consists in producing dynamically the resonance R which
is shown diagrammatically in Fig.~\ref{f2}. Then in the
$J/\psi\rightarrow\gamma \mathrm{VV}$ decay we naturally have a
process as shown in Fig.~\ref{f3} such that the part of the
$J/\psi\rightarrow\gamma\mathrm{R}$ amplitude is then given by the
process shown in Fig.~\ref{f4}, where the VV loop stands for the VV
propagator, or $G$ function, which appears in the scattering amplitude
for two vectors
\begin{equation}
 T=(1-\tilde{V}G)^{-1}\tilde{V}
\end{equation}
with $\tilde{V}$ the VV potential. In Fig.~\ref{f4} one also has the
couplings, $g_j$, of the resonance R to the different VV
intermediate states, which are calculated and tabulated in
Ref.~\cite{gengvec,Geng:2009gb}. Altogether the amplitude for
$J/\psi\rightarrow\gamma \mathrm{R}$ is proportional to
\begin{equation}\label{eq:mat}
 t_{J/\psi\rightarrow \gamma \mathrm{R}} \propto\sum_j w_j G_j g_j \ .
\end{equation}
The resonance decay vertices do not appear in this expression, since
they are irrelevant for the discussion below, which focuses on
inclusive observables.\footnote{ Although experimentally a resonance
$R$  is often identified in a specific decay channel, say $\pi\pi$
for the $f_0(1370)$,  the $J/\psi\rightarrow \gamma\mathrm{R}$ decay
is reconstructed by dividing by the branching ratio of R to this
channel. Thus, the whole inclusive $J/\psi\rightarrow\gamma
\mathrm{R}\rightarrow\gamma(\mathrm{R}\rightarrow\mathrm{all})$
decay is obtained and this is what we calculate by means of the
$t$-matrix of Eq.~(\ref{eq:mat}).} In Table I the values for the
$G_j$ and $g_j$ at the resonance peak obtained in
Ref.~\cite{gengvec,Geng:2009gb} are given.\footnote{We have given
only the real part of the $G_j$'s and
  $g_j$'s since their imaginary part is small for most cases. The
  uncertainties induced by using either the full complex value or only
  the real part are well within the range of uncertainties that we
  estimate below.}
Note that the main component of the $f_2(1270)$ and
$f_0(1370)$ ($f_2'(1525)$ and $f_0(1710)$) is the $\rho\rho$
($K^*{\bar K}^*$) pair, and the interference between the $\rho\rho$
and $K^*{\bar K}^*$ is constructive (destructive) for the
$f_2(1270)$ and $f_0(1370)$ ($f_2'(1525)$ and $f_0(1710)$).

\begin{table*}[htpb]
      \renewcommand{\arraystretch}{1.5}
     \setlength{\tabcolsep}{0.3cm}
\caption{$G_j$'s and $g_j$'s appearing in Eq.~(\ref{eq:mat}), with $j$ one of the coupled channels:
$\rho\rho$, $K^*\bar{K}^*$, $\omega\omega$, and $\phi\phi$. The $G_j$'s are in units of $10^{-3}$, and
the $g_j$'s are in units of MeV.
\label{table:sum}}
\begin{center}
\begin{tabular}{c|cc|cc|cc|cc}\hline\hline
R & \multicolumn{2}{c|}{$\rho\rho$} & \multicolumn{2}{c|}{$K^*\bar{K}^*$} & \multicolumn{2}{c|}{$\omega\omega$} & \multicolumn{2}{c}{$\phi\phi$}          \\\hline
& $G$ & $g$ & $G$ & $g$ & $G$ & $g$ & $G$ & $g$\\\hline
$f_2(1270)$ & $-4.37$ &$10889$ & $-2.34$ & $4733$ & $-4.97$ & $-440$ & $0.47$ & $-675$\\
$f'_2(1525)$ & $-8.33$ & $-2443$ & $-4.27$& $10121$ & $-9.62$ & $-2709$ & $-0.71$ & $-4615$\\
$f_0(1370)$ & $-8.09$ & $7920$ & $-4.14$ & $1208$ & $-9.11$ & $-39$ &$-0.63$ & $12$\\
$f_0(1710)$ & $-10.02$ & $-1030$ & $-7.81$ & $7124$ & $-11.15$ & $-1763$ & $-2.17$ & $-2493$ \\
 \hline\hline
    \end{tabular}
\end{center}
\end{table*}
\begin{table*}[htpb]
      \renewcommand{\arraystretch}{1.5}
     \setlength{\tabcolsep}{0.3cm}
\caption{Ratios of  $R_T=\Gamma_\mathrm{J/\psi\rightarrow \gamma f_2(1270)}/\Gamma_{J/\psi\rightarrow
\gamma f'_2(1525)}$ and $R_S=\Gamma_{J/\psi\rightarrow \gamma f_0(1370)}/\Gamma_{J/\psi\rightarrow\gamma
f_0(1710)}$ within the molecular model and the quark model
 in comparison with data~\cite{Amsler:2008zzb}.
\label{table:ratio}}
\begin{center}
\begin{tabular}{c|ccc}\hline\hline
 & Molecular picture & Quark model & Data\\\hline
$R_T$ & $2\pm 1$ & $2.2$ & $3.18^{+0.58}_{-0.64}$ \\
$R_S$ & $1.2\pm 0.3$ & $2.2-2.5$ & \\
$R_S/R_T$ & $0.6\pm 0.1$ & $1-1.1$ & \\
 \hline\hline
    \end{tabular}
\end{center}
\end{table*}

For inclusive resonance production
the partial decay widths are given
by
\begin{equation}
 \Gamma=\frac{1}{8\pi}\frac{1}{M_{J/\psi}^2}|t|^2 q
\end{equation}
where $q$ is the photon momentum in the $J/\psi$ rest system. For simplicity we here assume that
the final phase space can be calculated with the nominal resonance
mass. A more refined study would call for a proper folding with
 the resonance mass distribution (recall that this folding for the intermediate VV states is already done in
Ref.~\cite{gengvec,Geng:2009gb}).
 While this is relevant when one has decays close to the threshold of the
final state, in the present case there is plenty of phase space for
the $J/\psi\rightarrow\gamma\mathrm{R}$ decay and the folding barely
changes the results calculated at the central value of the resonance
mass distribution.

The most difficult part of this study is the determination of the
uncertainties. Since there is no proper effective field theory
underlying this study, we need to estimate the uncertainties of the
model via the uncertainties of the input quantities.  The model used
has 5 subtraction constants as parameters in the strangeness-zero
channels~\cite{gengvec,Geng:2009gb}.  They were all demanded to take
natural values, however, two of them were tuned a bit to get the
masses of the tensor--mesons in agreement with data.  In order to
determine the uncertainties of the result we now vary all parameters
independently: we produced a large number of results emerging from
calculations with different subtraction constants under the
constraint that the masses of the tensor mesons still are
reproduced. In addition the vector coupling constant was varied
within its 10 \% uncertainty. From this study a 95 \% confidence
level could be determined.  The corresponding results that we get
within the molecular picture are shown in the first column in Table
II and compared with the experimental data when
available~\cite{Amsler:2008zzb}. Please note that the uncertainty of
the super--ratio $R_S/R_T$ is considerably lower, since the
uncertainties in $R_S$ and $R_T$ are correlated.

We can see that we obtain a band of values perfectly compatible with
the experimental data for the ratio of rates of $J/\psi$ to $\gamma
f_2(1270)$ and $\gamma f'_2(1525)$. We get a central value of 2 for
this ratio.
 As we will show below, the quark model prediction for this quantity,
 assuming that the resonances belong to the same flavor nonet,
 is similar.

In addition to the tensor channel the model also makes predictions
for the ratio of the decay rates to the two scalar states
$f_0(1370)$ and $f_0(1710)$, which are also dynamically generated in
the approach of Ref.~\cite{gengvec,Geng:2009gb}. The central value
obtained is about 1.2 for the ratio of the decay rates to $\gamma
f_0(1370)$ and $\gamma f_0(1710)$. One can trace this smaller value
now, with respect to the case of the tensors, to a less efficient
cancellation between $K^*\bar{K}^*$ and $\rho\rho$ in the case of
the $f_0(1710)$ compared to the $f_2'(1525)$, where the $\rho\rho$
coupling has also opposite sign to that of $K^*\bar{K}^*$. The SU(3)
flavor content of the tensor and scalar resonances is similar within
the model employed: $f_0(1370)$ and $f_2(1270)$ are mostly
$\rho\rho$ molecules, while $f_0(1710)$ and $f'_2(1525)$ are mostly
$K^*\bar{K}^*$ molecules.  Yet, the nontrivial dynamics of the
coupled channels and the potentials $\tilde{V}$ of the hidden-gauge
approach tend to produce a factor of two difference in the ratios of
the decay rates of $J/\psi$ into states with similar flavor
contents, although the uncertainties between the two ratios still
overlap.

However, since the uncertainty of the super-ratio $R_S/R_T$ is
smaller, the difference between 0.6 in the case of the molecular
picture and the value around unity in the quark model picture to be
discussed below, within their uncertainties, are quite distinct.
Certainly the measurement of the ratio of decays to the scalar
mesons should be a valuable piece of information to further test the
nature of the resonances under consideration.

In the case of scalar mesons the intermediate pseudoscalar pairs
(they are in L=0 for the scalar mesons) in the sum of
Eq.~(\ref{eq:mat}) could provide some contribution.  We find the
most extreme case for the $f_0(1370)$, where the combination of
$G_jg_j$ of Eq.~(\ref{eq:mat}) for $\pi\pi$ might be of the same
order of magnitude as for the $\rho\rho$ component, however, almost
purely imaginary, so that there is no interference with the
$\rho\rho$ contribution. One must then rely upon the weighs $w_j$ to
decide the relative size of the $\pi\pi$ and $\rho\rho$
contribution. The limited information from the PDG (see rates
$\Gamma_{135}$, $\Gamma_{148}$, $\Gamma_{171}$) indicate that the
rates for $J/\psi\rightarrow\gamma\rho\rho$ might be about one order
of magnitude larger than for $J/\psi\rightarrow\gamma\pi\pi$
~\cite{Amsler:2008zzb}, so one can induce a smaller contribution of
intermediate pions, but this limited experimental information should
translate into larger uncertainties in the $R_S$ ratio of Table
\ref{table:ratio}, of the order of an additional 10-20 \%.

Please note that for the decay modes  $J/\psi \to \phi (\omega)
f_2(1270)$, $J/\psi \to \phi (\omega) f'_2(1525)$ and $J/\psi \to
K^{*0}(890) \bar{K}^{*0}_2(1430)$, estimated within the same
molecular picture as used in this section, a good agreement with the
data was achieved~\cite{albertozou} --- see Fig.~\ref{f5}.
 In the next section we shall
discuss the corresponding predictions for all the channels mentioned
within an admittedly naive quark model.

\section{Flavor counting considerations}
It is worth mentioning that the ratio of rates, $R_T$, obtained in
Table II is related to the dominance of the strange components in
$f'_2(1525)$ and the nonstrange ones in $f_2(1270)$. In a simple
$q\bar{q}$ model for these states one assumes that they belong to a
nonet of tensor mesons, which includes the
$K^*_2(1430)$~\cite{physrep}.  Certainly, in  a quark model mixing
between the SU(3) singlet $(u\bar{u}+d\bar{d}+s\bar{s})/\sqrt{3}$
and SU(3) octet isoscalar $(u\bar{u}+d\bar{d}-2s\bar{s})/\sqrt{6}$
is allowed. Since the mass difference between the $f_2(1270)$ and
the $f_2'(1525)$ is approximately the same as that between the
$\omega$ and the $\phi$, we may assume ideal mixing between them,
i.e., the $f_2(1270)$ corresponds to $(u\bar{u}+d\bar{d})/\sqrt{2}$,
the $f_2'(1525)$ to $s\bar{s}$ (see footnote \footnote{This is supported by the
studies of the strong and radiative decays of the $f_2(1270)$ and
$f'_2(1525)$, see, e.g., Refs.~\cite{ Barnes:1996ff, Barnes:2002mu,
Li:2000zb,Anisovich:2002im, Giacosa:2005bw}.}). 
The SU(3) singlet combination is now given by
\begin{equation}
 S\sim u\bar{u}+d\bar{d}+s\bar{s}=\sqrt{2}\frac{1}{\sqrt{2}}{(u\bar{u}+d\bar{d})}+s\bar{s},
\end{equation}
which provides a branching ratio
\begin{equation}\label{eq:flavora}
 \frac{B_{J/\psi\rightarrow \gamma f_2(1270)}}{B_{J/\psi\rightarrow\gamma f'_2(1525)}}
=\left(\frac{\sqrt{2}}{1}\right)^2\frac{q_2}{q'_2}\approx 2.2,
\end{equation}
where $q_2$ and $q'_2$ are the momenta of the $f_2(1270)$ and
$f'_2(1525)$ in the $J/\psi$ rest frame. This number is comparable
to the one obtained in the molecular model discussed in the
previous section, where the $f_2(1270)$ is
mostly $\rho\rho$ and the $f'_2(1525)$ is mostly $K^*\bar{K}^*$.
A fully dynamical quark model calculation for the mentioned
transitions,
e.g. along the lines of Refs.~\cite{Barnes:1996ff,Barnes:2002mu},
where the strong decays of
the $f_2(1270)$ and the $f_2'(1525)$ are
 well-described in the $^3 P_0$ quark model,
 would be most welcome.

Analogous flavor counting arguments provide  values between 2.2 and 2.5 for
the ratio $R_S$ of Table
II, depending on the masses used for the
scalar states, which is considerably higher than the value found
within the molecular model.

\begin{figure}[htbp]
\begin{center}
\includegraphics[scale=1.0]{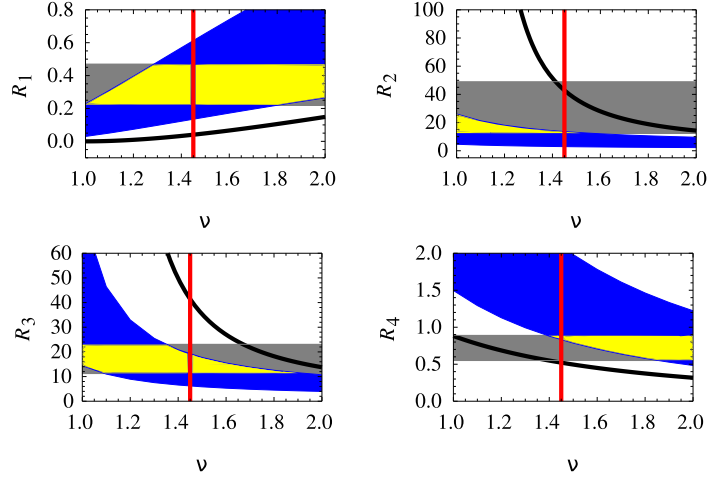}
\caption{(Color on line) Comparison  with data (gray shaded bands) of
$R_1$, $R_2$, $R_3$, and $R_4$  calculated with Eqs.~(14-17) within the quark model (solid black lines).
The vertical solid red lines indicate $\nu=1.45$
obtained in Ref.~\cite{albertozou}.
The results for the ratios of that reference, obtained using
the molecular model, are shown
by the blue shaded bands. The yellow regions indicate the overlap of
the results of the molecular model with the data.} \label{f5}
\end{center}
\end{figure}

There are similar transitions that can be studied within the
same framework, namely the ratios between the $J/\psi$ decays into
$\phi(\omega)f_2(1270)$, $\phi(\omega)f'_2(1525)$,
$K^{*0}\bar{K}^{*0}_2(1430)$. Within the molecular model
of the previous section those were studied in Ref.~\cite{albertozou}
--- the results of that reference, which describe the data well, although
with a sizable uncertainty, are indicated in Fig.~\ref{f5} as the blue shaded bands. It is
straightforward to estimate the same ratios also within the simple
quark model outlined above. Using the formulas obtained in Table I
of Ref.~\cite{albertozou} for the production of the $s\bar{s}$,
$\frac{1}{\sqrt{2}}(u\bar{u}+d\bar{d})$, and $s\bar{d}$ components,
together with $\phi$, $\omega$ or $K^{*0}$, one immediately finds
\begin{eqnarray}\label{R1234}
 R_1&=&\frac{\Gamma_{J/\psi\rightarrow\phi f_2(1270)}}{\Gamma_{J/\psi\rightarrow
\phi f'_2(1525)}}=2\left(\frac{\nu-1}{\nu+2}\right)^2\frac{q_2(\phi)}{q'_2(\phi)},\\
 R_2&=&\frac{\Gamma_{J/\psi\rightarrow\omega f_2(1270)}}{\Gamma_{J/\psi\rightarrow
\omega f'_2(1525)}}=\frac12\left(\frac{2\nu+1}{\nu-1}\right)^2\frac{q_2(\omega)}{q'_2(\omega)},\\
 R_3&=&\frac{\Gamma_{J/\psi\rightarrow\omega f_2(1270)}}{\Gamma_{J/\psi\rightarrow
\phi f_2(1270)}}=\frac12\left(\frac{2\nu+1}{\nu-1}\right)^2\frac{q_2(\omega)}{q_2(\phi)},\\
 R_4&=&\frac{\Gamma_{J/\psi\rightarrow K^{*0}\bar{K}^{*0}(1430)}}{\Gamma_{J/\psi\rightarrow
\omega f_2(1270)}}=\left(\frac{3}{2\nu+1}\right)^2\frac{q_2(K^{*0})}{q_2(\omega)},
\end{eqnarray}
where the $q_2(\mathrm{M})$'s and $q'_2(\mathrm{M})$'s are the
momentum of the meson M in the $J/\psi$ rest frame in the
corresponding decays, and $\nu$ measures the ratio of amplitudes for
producing simultaneously two singlets  and two octets of SU(3) in
the $J/\psi$ decay into $\phi$ ($\omega$)~\cite{albertozou}.

Before comparing the resulting ratios to the data it is useful
to estimate the allowed range for $\nu$.
For that purpose we may switch to the quantity
$\lambda_\phi$ defined in Ref.~\cite{ulfjose},   which is a measure of a subdominant
component in the $J/\psi$ decay to $\phi$ and a pair of $q\bar{q}$ prior
to hadronization as defined by
\begin{equation} J/\psi\rightarrow\phi[s\bar{s}+\lambda_\phi\frac{1}{\sqrt{2}}(u\bar{u}+d\bar{d})]\rightarrow
\phi MM,
\end{equation}
  and given in terms of $\nu$ by
\begin{equation}
 \lambda_\phi=\sqrt{2}\left(\frac{\nu-1}{\nu+2}\right) \ .
\end{equation}

\begin{figure}[t]
\begin{center}
\includegraphics[scale=1]{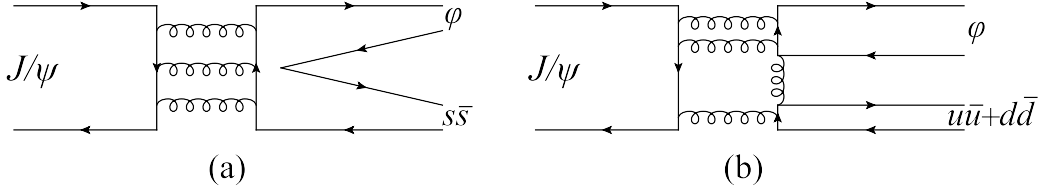}
\vglue-0mm\caption{$J/\psi$ decays into $\phi+q\bar q$ with one-OZI
suppression (a) and double-OZI suppression (b).} \label{fig:ozi}
\end{center}
\end{figure}
One can estimate the natural value of the OZI violation parameter
$\lambda_\phi$ using $N_c$ counting, with $N_c$ being the number of
colors. In Fig.~\ref{fig:ozi}, we show the mechanisms of the
$J/\psi$ decays into $\phi+q\bar q$ with single-OZI suppression and
double-OZI suppression, as shown by (a) and (b), respectively.
Taking into account that the strong coupling constant $g_s$ behaves
as $1/\sqrt{N_c}$, and counting the closed loops by changing the
gluon lines to double lines,  one can see that diagram (a) counts as
$N_c^0$, and (b) counts as $1/N_c$. Hence the double-OZI suppression
mechanism shown in (b) is suppressed by a factor of $1/N_c$ compared
with the single-OZI mechanism (a). Taking $N_c=3$, we get
$\lambda_\phi\sim0.33$. Correspondingly, $\nu\approx1.9$. Note that
this number provides only a crude estimate, and the value
$\nu=1.45$, found in Ref.~\cite{albertozou}, as well as those used
in Ref.~\cite{ulfjose} are roughly consistent with this estimate.

 Experimentally one finds a band of values
for each ratio: $R_1\approx0.22-0.47$,
$R_2\approx12.33-49.00$, $R_3\approx11.21-23.08$, and $R_4\approx
0.55-0.89$~\cite{albertozou}.
In Fig.~\ref{f5}, as the black solid line, the predictions from the naive quark model
for the ratios are shown as a function of $\nu$ in the parameter range
allowed. Also shown in the figure are the experimental data (gray shaded bands) as well as the predictions of Ref.~\cite{albertozou} (blue shaded bands).
 As can be seen from the figure, while the molecular model appears to
 be fully consistent with the data, there is no value for $\nu$ in the
 range allowed that brings all ratios in agreement with the data --- however,
 for $\nu=1.7$ or larger, $R_2$ and $R_3$ agree with the data, while $R_1$
 and $R_4$ are off only by two sigma, which one might still view as acceptable
 given the crudeness of the quark model used here.
The important message of Fig.~\ref{f5} is that the predictions of the molecular
model discussed here and the naive quark model are quite distinct.
Further experimental and theoretical analyses are necessary to really test the nature of these resonances
using the processes studied here.

\section{Further discussions}

As can be seen from the numbers listed in Table~\ref{table:ratio},
the central value of $R_T$ in the molecular model is similar to that
 in the quark model. However, they come from quite
different physical sources. In the quark model, $R_T\approx2.2$
comes mainly from the SU(3) flavor wave functions of the two tensor
mesons modulo small kinematic correction as seen from
Eq.~(\ref{eq:flavora}). In the molecular model, the result about 2
comes from the nontrivial interference pattern between the dominant
channel and the others. To be explicit, using the values of $G_i$
and $g_i$ given in Table~\ref{table:sum} and $w_i$ given in
Eq.~(\ref{eq:cg}), one gets $t_o/t_d=0.29$ for the $f_2(1270)$ and
$t_o/t_d=-0.07$ for the $f_2'(1525)$, where $t_d$ and $t_o$ denote
the decay amplitudes from the dominant component and the other
components, respectively. Note that
$\left(\frac{t_d[f_2(1270)]}{t_d[f'_2(1525)]}\right)^2\approx0.9$
and $(1.29/0.93)^2\approx2$. Contrary to the case of the tensor
mesons, in the molecular model, the amplitude from the dominant
component only gets a small correction from non-trivial interference
with the other components for both the $f_0(1370)$ and the
$f_0(1710)$. Moreover, the corrections for these two scalars are
similar, $t_o/t_d$ is 9\% for the $f_0(1370)$, and  6\% for the
$f_0(1710)$ (using the numbers given in Table \ref{table:sum}).  As
a result, the value of $R_S$ is approximately 1 in the molecular
model, which differs from that in the quark model.

 We must mention that the current version of the PDG
review~\cite{Amsler:2008zzb} quotes the observation of the
$f_0(1710)$ in three $J/\psi$ radiative decay modes, i.e.,
$J/\psi\rightarrow\gamma f_0(1710)\rightarrow \gamma K\bar{K}$,
$J/\psi\rightarrow \gamma f_0(1710)\rightarrow \gamma \omega\omega$,
and $J/\psi\rightarrow \gamma f_0(1710)\rightarrow\gamma \pi\pi$,
while no clear $f_0(1370)$ has been seen in these data. This, at
first sight, seems to contradict the results in both the molecular
model ($R_S=1.2\pm0.3$) and the quark model ($R_S\approx2.2-2.5$).
However, this might be a consequence of the decay channels
experimentally studied. For instance, in the molecular model used
here the $f_0(1370)$  couples only  very weakly to both $K\bar{K}$
and $\omega\omega$. Thus, the non-observation of this state in the
$\gamma K\bar{K}$ and $\gamma \omega\omega$ decay modes of the
$J/\psi$ decays should be of no surprise in that model. On the other
hand, since in that model the $f_0(1370)$ couples much more strongly
to $\pi\pi$ than the $f_0(1710)$ does, it seems mysterious that in
the $J/\psi\rightarrow \gamma\pi\pi$ data~\cite{Ablikim:2006db} only
the $f_0(1710)$ is seen but not the $f_0(1370)$. A possible
explanation is that the scalar state peaking at
$1765^{+4}_{-3}\pm13$ MeV with a width of $145\pm8\pm69$
MeV~\cite{Ablikim:2006db} might not be the same as the $f_0(1710)$
observed in the $J/\psi\rightarrow\omega\pi\pi(K\bar{K})$. This
conjecture is supported by the fact that the $f_0(1710)$ branching
fractions to $\pi\pi$ and $K\bar{K}$ obtained from
$J/\psi\rightarrow\gamma\pi\pi(K\bar{K})$
data~\cite{Ablikim:2006db,Bai:2003ww} and $J/\psi\rightarrow \omega
\pi\pi (K\bar{K})$ data~\cite{Ablikim:2004qna,Ablikim:2004st},
$0.41^{+0.11}_{-0.17}$ and $<0.11$ at 95\% confidence level,
respectively,  are not consistent with each other, which indicates
that the states observed in these two sets of data might indeed be
different. On the other hand, the $f_0(1710)$ in the molecular
picture has a branching ratio $\Gamma_{\pi\pi}/\Gamma_{K\bar{K}}$ at
a few percentage level~\cite{gengvec,Geng:2009gb}, consistent with
the $f_0(1710)$ observed in $J/\psi\rightarrow
\omega\pi\pi(K\bar{K})$~\cite{Ablikim:2004qna,Ablikim:2004st}.

The new analysis of the BES II
$J/\psi\rightarrow\gamma(\pi\pi\pi\pi)$ data~\cite{Bugg:2009ch},
which updates the analyses of the Mark III~\cite{Bugg:1995jq} and
BES data~\cite{Bai:1999mm}, claims that the scalar state around
$1.7\sim1.8$ GeV is the $f_0(1790)$, which is also seen in
$J/\psi\rightarrow\phi\pi\pi$~\cite{Ablikim:2004wn}. The $f_0(1370)$
is also definitely needed to fit the data, while no $f_0(1710)$ is
necessary. The observations are consistent with the molecular
picture for the $f_0(1710)$ and $f_0(1370)$. Indeed, since in the molecular
model the ratio $R_S=\frac{\Gamma_{J/\psi\rightarrow\gamma f_0(1370)}}{\Gamma_{J/\psi\rightarrow \gamma f_0(1710)}}$ is about 1 and the $f_0(1370)$ couples more strongly to
$\rho\rho$ than the $f_0(1710)$ does, one would naively expect to
see more $f_0(1370)$ than $f_0(1710)$ in $J/\psi\rightarrow\gamma
(\pi\pi\pi\pi)$.

The study of scalar mesons between 1 and 2 GeV has always been
complicated by possible mixing with nearby glueballs, see e.g.
Ref.~\cite{Close:2005vf} and references therein. Things might become
more subtle if there exists an extra scalar state, the $f_0(1790)$,
in addition to the $f_0(1370)$, $f_0(1500)$, and $f_0(1710)$.  In
the molecular picture of Ref.~\cite{gengvec,Geng:2009gb}, possible
mixing of glueballs with the $f_0(1370)$ and $f_0(1710)$ is not
considered, which is in line with the study of
Ref.~\cite{Close:2005vf}. In that paper, it is suggested that the
$f_0(1500)$ has a large glueball component while the $f_0(1370)$ and
$f_0(1710)$ have relatively small glueball components. Certainly,
one should keep in mind that the $f_0(1370)$ and $f_0(1710)$ in the
molecular model are dynamically generated states from vector
meson-vector meson interactions while those in
Ref.~\cite{Close:2005vf} are considered as mainly $q\bar{q}$ states,
and hence, one should not expect the same mixing pattern in these
two pictures.

\section{Summary and conclusions}
We have carried out an evaluation of the ratios of the rates for the
$J/\psi \to \gamma f_2(1270)$, $J/\psi \to \gamma f'_2(1525)$,
$J/\psi \to \gamma f_0(1370)$, and $J/\psi \to \gamma f_0(1710)$
decays.  The ratios were estimated either using a molecular model,
where all the mentioned states emerge as bound states or resonances
of two vector mesons, as proposed in
Ref.~\cite{gengvec,Geng:2009gb}, or using a simple quark model.  The
results obtained for the ratio of the rates in the production of the
tensor mesons is in good agreement with experiment for both
approaches. We then make predictions for the ratio of rates for
$J/\psi$ decays into the scalar mesons, which could be tested in the
future.

We also compare the predictions of the molecular model for
 $J/\psi$ decays into $\phi$, $\omega$, $K^{*0}$ and together with the
mentioned resonances~\cite{albertozou} to those from the same quark
model. Here the molecular picture can describe the decay
ratios well, while the results from the simplified quark model
are consistent with the data only within two sigma. Improved data are
desirable to draw more firm conclusions.

In addition, a corresponding dynamical quark model calculation ---
including estimates of uncertainties, which was not possible in the
simplified quark model used here --- would be very important for the
$J/\psi$ decays discussed in this paper in order to better understand
the nature of the $f_2(1270)$, $f'_2(1525)$, $f_0(1710)$, and
$f_0(1370)$.

\section*{Acknowledgments}
L.S.G and E.O would like to thank Alberto Mart\'{i}nez for useful
discussions. We also thank Ulf-G. Mei{\ss}ner for a
careful reading of the manuscript and for his useful comments. This
work is partly supported by DGICYT contract number FIS2006-03438. We
acknowledge the support of the European Community-Research
Infrastructure Integrating Activity ``Study of Strongly Interacting
Matter" (acronym HadronPhysics2, Grant Agreement n. 227431) under
the Seventh Framework Program of EU. L.S.G. acknowledges support
from the MICINN in the Program ``Juan de la Cierva.'' F.K.G. and
C.H. also acknowledge the support of the Helm\-holtz Association
through funds provided to the virtual institute ``Spin and strong
QCD'' (VH-VI-231) and by the DFG (SFB/TR 16, ``Subnuclear Structure
of Matter''). B.S.Z. acknowledges support from the National Natural Science Foundation of China.

\end{document}